\documentclass[prl,twocolumn,showpacs]{revtex4}

\usepackage{graphicx}
\usepackage{dcolumn}
\usepackage{bm}

\begin{document}

\title{Conversion of phase-slip lines into elementary resistive domains in a
current-carrying superconducting thin film}

\author{E.~V.~Il'ichev}
\altaffiliation [ Now at ] { Institute for Physical High
Technology, P.O. Box 100239, D-07702 Jena, Germany}
\author{V.~I.~Kuznetsov}
\email[ Electronic address: ] {kvi@ipmt-hpm.ac.ru}
\author{V.~A.~Tulin}
\affiliation{Institute of Microelectronics Technology and High Purity Materials,
Russian Academy of Sciences, 142432 Chernogolovka, Moskow Region, Russia}

\begin{abstract}
A nonmonotonic dependence of the differential resistance of the first step
of a phase-slip line on the heat-reservoir temperature near $T_{c}$ has been
observed experimentally for the first time in a superconducting tin film.
This behavior is interpreted as a conversion of the phase-slip line into an
elementary resistive domain as the result of a deviation from the isothermal
conditions.
\end{abstract}

\maketitle

The mechanisms by which superconductivity is destroyed when a transport
current is passed through a thin film are basically local mechanisms. Narrow
films, whose width $w$ is smaller than the coherence length $\xi$, stratify into
so-called phase-slip centers in the process \cite{ivlev}. In wide samples, an
increase in the current is accompanied by a penetration of flux vortices
from the edges of the film, near defects and irregularities. This effect
results in a dissipation of energy and a local heating. Resistive regions or
normal regions ( if the temperature rises to a value $T_{m} > T_{c}$ ) which are
nonequilibrium and nonisothermal but nevertheless localized form \cite{ivanchenko,yu}.
The current-voltage characteristic of such a film is stepped and has a
substantial thermal hysteresis. If the heating is moderate ( $ T_ {m} \approx T_ {0} $,
where $T_{0}$ is the heat-reservoir temperature),
the film is unstable with respect to the formation of nonequilibrium, nearly
isothermal regions of phase-slip lines: 2D analogs of phase-slip centers
\cite{vol,volotsk}. The current-voltage characteristic of samples containing phase-slip
lines is free of hysteresis and is stepped. It consists of a series of
linear sections with differential resistances $R_{n} = n R_{0}$, $n =1,2,...$.
The resistance $R_{0}$ is determined by the resistance of the film in its normal
state over a distance $ 2l_ {E} $, where $ l_ {E} $ is the depth to which an electric field
penetrates into the superconductor \cite{volotsk}.

When elementary resistive domains \cite{ivanch} form, the current-voltage characteristic may
have qualitatively the same shape as for phase-slip lines. There is the distinction that the
depth to which the electric field penetrates upon the formation of the elementary resistive
domains is different, and it depends on the heating and the heat transfer. In addition,
a thermal hysteresis appears on the current-voltage characteristic in the
presence of an elementary resistive domain \cite{ivanch}. The formation of elementary
resistive domains experimentally is improbable. If the heat transfer is
poor, one usually observes a coalescence of the elementary resistive domains
into large domains. The smallest resistive domains which have been observed
\cite{ivanch,medvedev} consist of six to ten elementary domains.

 \begin{figure}

 \includegraphics[width = 1.0\linewidth]{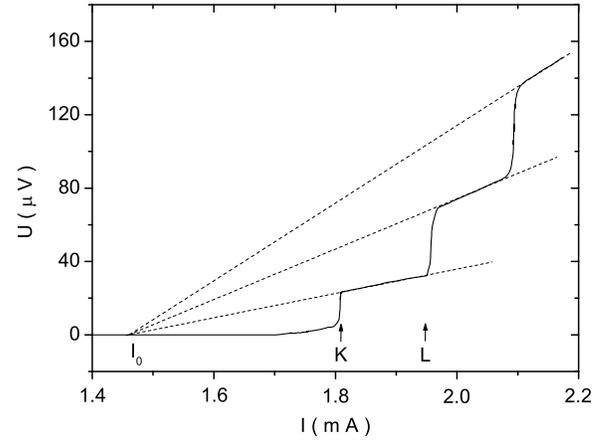}

\caption{\label{fig 1.} Current-voltage characteristic of a sample at
$T = 3.86 K $; $I_{0}$ is the excess current.}

\end{figure}

Phase-slip lines exist in only a narrow temperature interval \cite{volotsk}, so neither
the temperature dependence of the lines nor the changes caused in them by a
deviation from isothermal conditions have previously been studied. In the
present letter we are reporting a study of these questions.

 The preparation of the samples consisted of the following steps: (1) vacuum
deposition of tin to a thickness $ d  \approx 1000 $ {\AA} on silicon substrates; (2)
photolithography to fabricate strips with a width $ w = 70$ $ \mu$m and a
length $l = 2$ mm; (3) the
formation of a channel $ \approx 1$ $\mu$m wide and
$ \approx 200$ {\AA} deep running across the strip, by
electron-beam lithography and ion-beam etching. The lateral boundaries of
the strip were modulated with an amplitude and a period of a few microns to
facilitate the penetration of vortices. The channel served the same purpose.
Experiments revealed that phase-slip lines could be observed even without
the channel, but they were not as apparent in that case.

 We recorded current-voltage characteristics of the samples, finding the
temperature from the helium vapor pressure. Figure 1 shows a typical
characteristic; it is what we would expect for phase-slip lines \cite{volotsk}. There
are linear regions of substantial width along the current scale; up to 20
distinct steps are observed. The resistance of the $N$th step is $N \times R_{0}$, where
$R_{0}$ is the resistance of the first step. As the temperature
is lowered, this stepped structure persists, but a hysteresis appears.

We calculated the resistance of region \textit{KL} in Fig. 1 from the slope of step
$R_{0}$, and we calculated the difference between the resistances of the second
and first steps, $R_{1} - R_{0}$, as functions of the reservoir temperature $T_{0}$.
The curves of $ R_ {0} (T_ {0}) $ and
$R_ {1}(T) - R_ {0}(T_ {0}) $ are not monotonic (Fig. 2). In interval $EC$,
the points of the two plots coincide, and the positions of these points remain essentially
unchanged as the temperature is cycled. Region $CA$ has a significant scatter in
resistances (Fig. 2).

\begin{figure}

\includegraphics[width = 1.0\linewidth]{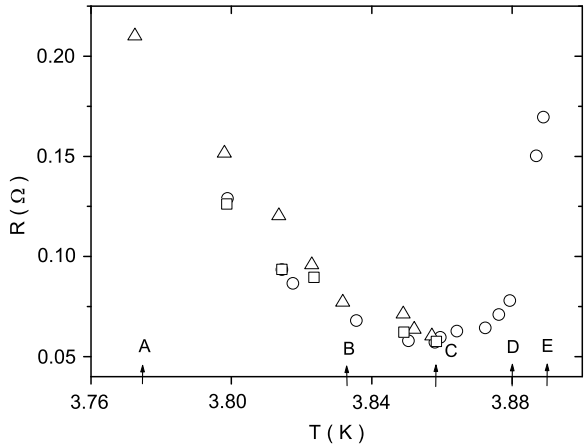}

\caption{\label{fig 2.} Temperature dependence of the average differential resistance.
${\rm O}$, $\Delta $ - resistance of the first step, for two measurement cycles;
open squares - difference between the resistances of the second and first
steps. $T_{c}=3.89$ K.}

\end{figure}

In region $ED$, which is close to $ T_ { c}$, we do not yet have any phase-slip lines.
In region $DC$ there is a multitude of phase-slip lines.

For a quantitative analysis we use the expression ( \cite{volotsk}, for example)

\begin{equation} R_{0} = 2 \rho_{n}l_{E}/wd,
\label{1}
 \end{equation}

 \noindent where $ \rho_{n} $ is the resistivity of the film in the normal state, and

\begin{equation} l_{E} = \lambda_{E}(1 - \tau)^{-1 / 4}.
\label{2}
 \end{equation}

 \noindent Here $\tau$ = $T$ / $T_{c}$ and we assume $T =$ $T_{ 0}$, since
 there is essentially no heating in interval $DC$; we thus have

\begin{equation} \lambda_{E} = (v_{F} l \tau_{å} / 3)^{1 / 2}.
\end{equation}

\noindent Here $v_{F} = 7 \times 10^{7}$ cm/s is the Fermi velocity, and
$\tau_{E} \approx 3 \times 10^ {-10} $ s is the energy relaxation time
for tin. The mean free path $l$ is usually found from the known value of $\rho_{n}l$.
However, there is a substantial scatter in the values of this quantity in the literature,
so we use the electron specific heat \cite{faber} $\gamma = 1030$ $erg/(deg^{2} cm^{3})$.
We also use the expressions $\kappa = 7.5 \times 10^{3} \rho_{n} \gamma^{1 / 2}$
and $\kappa = 0.75 \times \lambda_{L}(0)/l$, for the Ginzburg-Landau parameter \cite{gen},
where $\lambda_{L}(0)$ is the London depth at $T = 0$.
We thus find $\rho_{n} l \approx 1.6 \times 10^ { - 11}$ $\Omega$ $cm^ { 2}$. Using the
experimental value $\rho_{n} \approx 3 \times 10^ { - 6}$ $\Omega$ $cm$ for
our samples, we find $l \approx  500$ {\AA}.
Correspondingly, we have $\lambda_{E} \approx 1.9 $ $\mu$m. Then we find

\begin{equation} R_{0} = R(0)( 1-\tau)^{-1 / 4}.
 \end{equation}

\noindent Here $ R(0) = 2 \rho_{n} \lambda_ {E} / wd$.
Calculations yield $R(0) \approx 1.7 \times 10^{ - 2}$ $\Omega$.

\begin{figure}

\includegraphics [width = 1.0\linewidth]{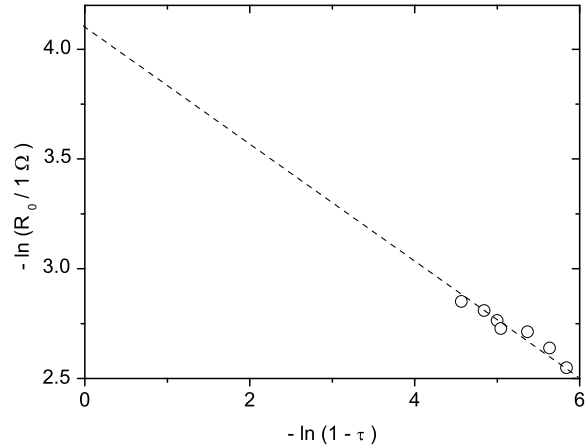}

\caption{\label{fig 3.} Temperature dependence of the resistance of a phase-slip line (for
interval $CD$ in Fig. 2).}

\end{figure}

Figure 3 shows a plot of $R = R_{0}/$(1$\Omega$ ) (i.e., a dimensionless quantity)
versus $ ( 1 - \tau ) $ in logarithmic scale for temperature interval $DC$. The experimental
points conform to a straight line with a slope
$\kappa \approx  - 0.26 $. A continuation of this straight line intersects the resistance
axis at the point $ln$ $R(0) \approx - 4.1$, from which we find
$R(0) \approx 1.7 \times 10^{ - 2} $ $\Omega$.
The experimental results thus not only confirm functional dependence (4) but also agree
quantitatively with it.

We know that resistive formations have two characteristic
parameters: $l_ {E} $, the penetration depth of the electric
field, and $ \lambda_ {T} $, the length scale of the temperature
decay with distance from the center of the domain
\cite{yu,ivanch}. The dependence of $l_ {E} $ on the parameters of
a sample was found in \cite{yu}, but that dependence does not
apply to our films, which have a good heat transfer. We will
accordingly restrict our explanation of the behavior of the
average differential resistance in region $CA$ (Fig. 2) to a few
qualitative comments. It is logical to suggest that the phase-slip
lines become nonisothermal here and that thermal processes play a
governing role. Specifically, (1) the hysteresis increase in
interval $BA$; (2) the steps on the current-voltage characteristic
become noticeably nonlinear; (3) the resistance of the first step
increases with decreasing temperature; and (4) estimates of $
\lambda_ {T} $ for the case of a slight heating ($T$ - $T_{0}$
$\ll$ $T_{0}$) in region $CB$ show that $\lambda_ {T} $ is on the
order of $l_ {E} $. This result means that deviations from an
isothermal situation have a substantial influence on the size of
the resistive formations. For estimates of $ \lambda_ { T} $ here
we used the expression $ \lambda_ {T} = (kd / H)^{1 / 2}$, where
$k \approx  3 \times 10^{ - 2}$ $W /(deg$ $cm)$ is the thermal
conductivity found from the Wiedemann-Franz law, and $H  \approx
2$ $W/(cm^{2}$ $deg$) is the heat-transfer coefficient from
\cite{neeper}.

In region $CB$ we are thus witnessing a transition from a phase-slip line to
elementary resistive domains. In interval $BA$, thermal processes are
determining the behavior of the current-voltage characteristic. Here the resistance and
the current at which the first step appears depend on the temperature in
roughly the same way; specifically, they are proportional to $ (1 - \tau )$.
It follows that the average differential resistance is linear in the current.
The increase in the resistance in this temperature interval is probably due
to a growth of the domain, whose size is proportional to the current at which it appears.


\begin{references}

\bibitem{ivlev}
B.~N.~Ivlev and N.~B.~Kopnin, Usp. Fiz. Nauk {\bf 142}, 435 (1984)
[Sov. Phys. Usp. {\bf 27}, 206 (1984)].

\bibitem{ivanchenko}
Yu.~M.~Ivanchenko, P.~N.~Mikheenko, and V.~F.~Khirnyi, Zh. Eksp. Teor. Fiz.
{\bf 80}, 171 (1981) [Sov. Phys. JETP {\bf 53}, 86 (1981)].

\bibitem{yu}
Yu.~M.~Ivanchenko and P.~N.~Mikheenko, Zh. Eksp. Teor. Fiz.
{\bf 82},488 (1982) [Sov. Phys. JETP {\bf 55}, 281 (1982)].

\bibitem{vol}
V.~G.~Volotskaya, I.~M.~Dmitrenko, L.~E.~Musienko, and A.~G.~Sivakov, Fiz. Nizk. Temp.
{\bf 7}, 383 (1981) [Sov. J. Low Temp. Phys. {\bf 7}, 188 (1981)].

\bibitem{volotsk}
V.~G.~Volotskaya, I.~M.~Dmitrenko, L.~E.~Musienko, and A.~G.~Sivakov, Fiz. Nizk. Temp.
{\bf 10}, 347 (1984) [Sov. J. Low Temp. Phys. {\bf 10}, 179 (1984)].

\bibitem{ivanch}
Yu.~M.~Ivanchenko and P.~N.~Mikheenko, Zh. Eksp. Teor. Fiz. {\bf 83}, 684 (1982)
[Sov. Phys. JETP {\bf 56}, 380 (1982)].

\bibitem{medvedev}
Yu.~V.~Medvedev and V.~F.~Khirnyi, Fiz. Tverd. Tela (Leningrad) {\bf 26}, 1163 (1984)
[Sov. Phys. Solid State {\bf 26}, 705 (1984)].

\bibitem{faber}
Faber and A.~B.~Pippard, Proc. R. Soc. London {\bf 231}, 336 (1955).

\bibitem{gen}
P.~G.~de~Gennes, \textit{Superconductivity of Metals and Alloys}, Benjamin, Inc., New York, 1966.

\bibitem{neeper}
D.~A.~Neeper and J.~R.~Dillinger, Phys. Rev. {\bf A 135}, 1028 (1964).

\end{references}
\end{document}